\documentclass[aip,amsmath,amssymb,reprint]{revtex4-1}

\usepackage{graphicx}
\usepackage{dcolumn}
\usepackage{bm}
\usepackage[utf8]{inputenc}
\usepackage[T1]{fontenc}
\usepackage{mathptmx}
\usepackage{etoolbox}
\usepackage{comment}
\usepackage{siunitx}
\usepackage{makecell}
\usepackage{upgreek}
\usepackage{amsmath,amssymb}
\usepackage{xcolor}
\DeclareSIUnit\angstrom{\text {Å}}
\makeatletter
\def\@email#1#2{
 \endgroup
 \patchcmd{\titleblock@produce}
  {\frontmatter@RRAPformat}
  {\frontmatter@RRAPformat{\produce@RRAP{*#1\href{mailto:#2}{#2}}}\frontmatter@RRAPformat}
  {}{}
}
\makeatother
\begin{document}

\preprint{AIP/123-QED}

\title[]{Spin-orbit torques and magnetization switching in Gd/Fe multilayers generated by current injection in NiCu alloys}
\author{Federica Nasr}
\author{Federico Binda}
\author{Charles-Henri Lambert}
\author{Giacomo Sala}
\author{Paul Noël}
\author{Pietro Gambardella}
\affiliation{Department of Materials, ETH Zurich, CH-8093 Zurich, Switzerland}
\email[]{Authors to whom correspondence should be addressed: federica.nasr@mat.ethz.ch, pietro.gambardella@mat.ethz.ch}
\date{29 December 2023}

\begin{abstract}
Light transition metals have recently emerged as a sustainable material class for efficient spin-charge interconversion. We report measurements of current-induced spin-orbit torques generated by Ni$_{1-x}$Cu$_{x}$ alloys in perpendicularly-magnetized ferrimagnetic Gd/Fe multilayers. We show that the spin-orbit torque efficiency of Ni$_{1-x}$Cu$_{x}$ increases with the Ni/Cu atomic ratio, reaching values comparable to those of Pt for Ni$_{55}$Cu$_{45}$. Furthermore, we demonstrate magnetization switching of a 20-nm-thick Gd/Fe multilayer with a threshold current that decreases with increasing Ni concentration, similar to the spin-orbit torque efficiency. Our findings show that Ni$_{1-x}$Cu$_{x}-$based magnetic heterostructures allow for efficient control of the magnetization by electric currents.
\end{abstract}

\maketitle

Materials that allow for converting charge currents into spin currents are used to generate spin-orbit torques (SOTs) in different classes of magnetic systems and devices.\cite{ManchonRMP2019} SOTs enable the energy-efficient manipulation of the magnetization in ferromagnets,\cite{MironNat2011, LiuScience2012, GrimaldiNatNanotec2020} ferrimagnets,\cite{FinleyPRA2016, CarettaNatNanotech2018, PhamPRA2018, SalaAMI2022} and antiferromagnets,\cite{OlejnikSciAdv2018, TsaiNat2020, KrishnaswamyPRA2022} with applications in magnetic random access memories,\cite{KrizakovaJMMM2022} racetrack memories,\cite{BlasingProcIEEE2020} logic devices,\cite{LuoNat2020} and spin-torque nano-oscillators.\cite{DemidovPR2017, ZahedinejadNatMater2022} SOTs have been extensively studied in heavy metal/ferromagnet bilayers,\cite{MironNat2011, LiuScience2012, PaiAPL2012, GarelloNatNanotec2013, KimNatMater2013, ManchonRMP2019, IshikuroPRB2019, MironNatMat2010} where the charge-to-spin conversion originates from the interplay of bulk and interfacial spin-orbit coupling, as exemplified by the spin Hall \cite{SinovaRMP2015} and Rashba-Edelstein effects.\cite{BihlmayerNRP2022} The two figures of merit describing the charge-to-spin conversion efficiency are the spin Hall angle and the spin Hall conductivity. The former is defined as the ratio of the transverse spin current generated in the spin injector to the applied charge current, whereas the latter is the ratio of the spin current to the external electric field driving the charge current.\cite{ManchonRMP2019, NguyenPRL2016}

A large spin Hall angle ($\ge 0.1$) is usually observed in the \emph{5d} transition metals, such as Ta, W, and Pt, owing to their strong atomic spin-orbit coupling.\cite{SinovaRMP2015} Whether light metals can also show large charge-to-spin conversion is a question of both fundamental and practical impact on developing new and low-cost spin injector materials.
In this context, \emph{3d} transition metals and their alloys have recently attracted interest because they present appreciable spin Hall angles despite their weak spin-orbit coupling.\cite{MiaoPRL2013, DuPRB2014, QuPRB2015, AnNatCommun2016, WangSciRep2017, BaekNatMat2018, HibinoPRA2020} The interest in light metals has further surged following the theoretical prediction of a large orbital Hall conductivity in the nonmagnetic \emph{3d} elements,\cite{KontaniPRL2009, JoPRB2018} which can lead to the generation of strong SOTs in adjacent ferromagnetic layers.\cite{GoPRS2020,LeeNatCommun2021,LeeCommunPhys2021,SalaPRR2022} 

For transition metals, besides the spin-orbit coupling, the \emph{d}-orbital occupation plays an important role in determining the magnitude and sign of the spin Hall angle.\cite{SinovaRMP2015, KontaniPRL2009} Ni is particularly interesting in this respect because it has the same valency and a similar band structure as Pt. The calculated spin Hall conductivity of Ni is $\sim 1.2\times10^5~\SI{}{\ohm^{-1}\m^{-1}}$, about half of Pt, whereas the orbital Hall conductivities of Ni and Pt are very similar.\cite{JoPRB2018} Extrinsic spin scattering effects due to alloying of different metallic elements can further enhance the charge-spin interconversion efficiency,\cite{ZhuAPR2021, HuAQT2020, ChiAPLMater2021, WangAdvMater2022} as shown, e.g., for Cu$_{1-x}$Bi$_{x}$ and Au$_{1-x}$Pt$_{x}$.\cite{NiimiPRL2012, ZhuPRA2018} In light of these observations, experimental work has focused on Ni$_x$Cu$_{1-x}$ alloys, in which the introduction of Cu reduces the Curie temperature relative to pure Ni, making the alloy paramagnetic at room temperature for $x \leq 70\%$.\cite{Bozorth1993, AhernPRSLA1958} 
Despite the alloying, the band structure of Ni$_x$Cu$_{1-x}$ remains similar to that of paramagnetic Ni, suggesting a significant intrinsic spin Hall effect regardless of chemical disorder.\cite{KellerPRB2019} Recent experiments have reported large effective spin Hall angles comparable to 5$d$ metals in Permalloy/Ni$_{60}$Cu$_{40}$, CoFeB/Cu/Ni$_x$Cu$_{1-x}$, and YIG/Ni$_{80}$Cu$_{20}$ bilayers.\cite{KellerPRB2019, VarottoPRL2020, WuPRL2022} 
Additionally, spin-to-charge conversion in CoFeB/Ni$_{70}$Cu$_{30}$ was shown to induce \SI{}{\tera\hertz} emission upon fs laser irradiation with intensity up to half that of CoFeB/Pt.\cite{ChengPRB2022}

These studies show that Ni$_{x}$Cu$_{1-x}$ alloys constitute an interesting material system for spin-charge interconversion and the realization of spin current detectors.
Here, we investigate Ni$_{x}$Cu$_{1-x}$ alloys as spin injectors for generating SOTs in perpendicularly magnetized Ni$_{x}$Cu$_{1-x}$/Ti/[Gd/Fe]\textsubscript{30} heterostructures.
We select three different compositions ($x = 40,\,50,\,55\%$) for which Ni$_{x}$Cu$_{1-x}$ is paramagnetic at room temperature and report composition-dependent SOT efficiencies, reaching values of up to 0.06 for $x = 55\%$, which corresponds to a spin Hall conductivity of $(1.04\pm0.07)\times10^5~\SI{}{\ohm^{-1}\meter^{-1}}$ without taking into account corrections due to spin memory loss and interface transparency.
In addition, we demonstrate the efficient magnetization switching capability of these systems, including a fourth alloy's composition $x = 70\%$ at the ferromagnetic phase transition. Our results show that Ni$_{x}$Cu$_{1-x}$ alloys are efficient spin current generators that can be used to achieve electrical control of the magnetization in spintronic devices.

\begin{figure}[t!]
\centering
\includegraphics[width=8.6cm,keepaspectratio]{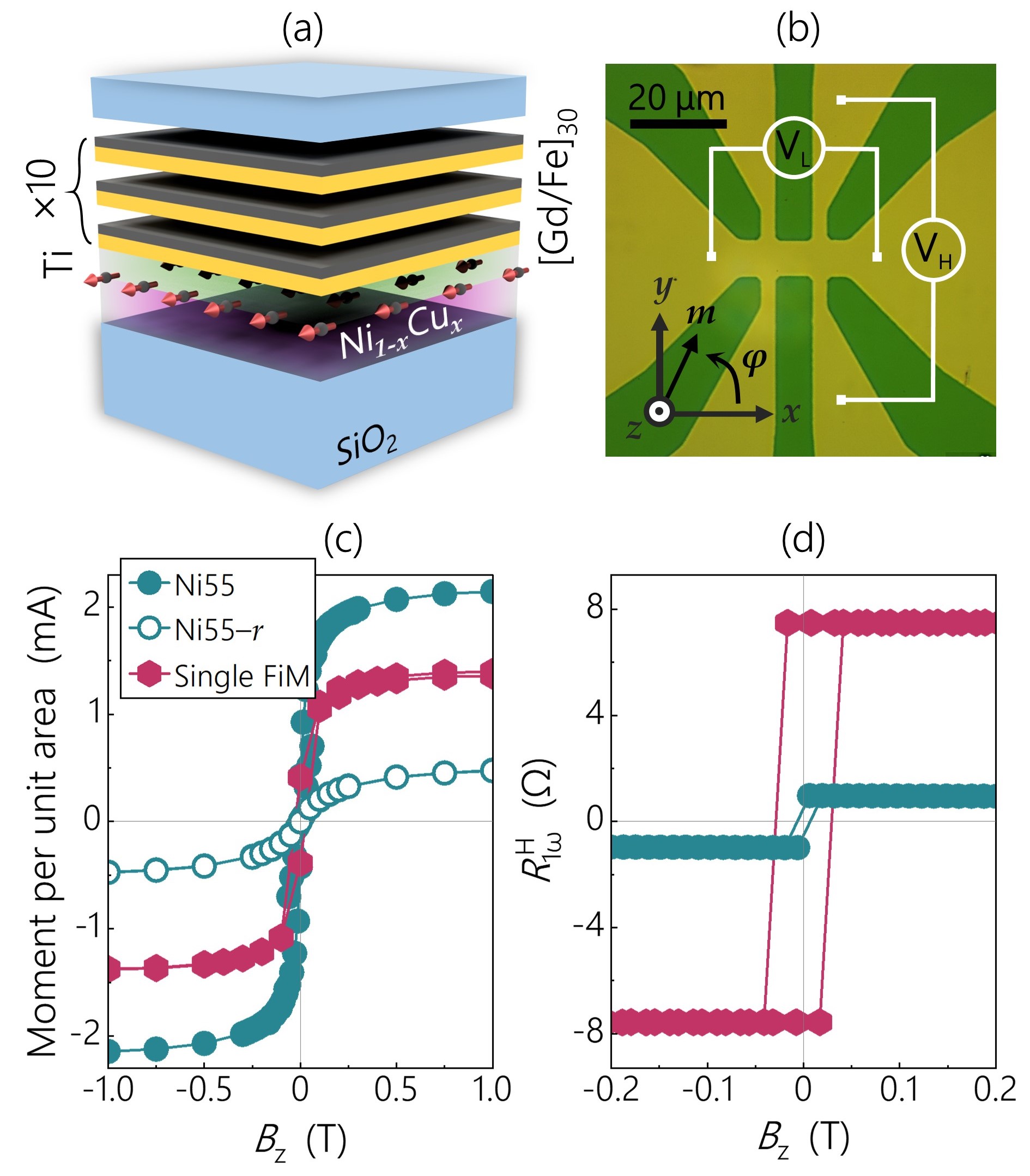}
\caption{\label{fig1} (a) Schematic of the Ni$x$ heterostructure. 
(b) Optical image of a Hall bar device with superimposed electrical connections and coordinate system.     
(c) Magnetic moment per unit area as a function of out-of-plane magnetic field for Ni55 (green dots), its reference Ni55$-r$ (open circles), and the single FiM sample (purple hexagons).
(d) Anomalous Hall resistance as a function of out-of-plane magnetic field for Ni55 (green dots) and the single FiM (purple hexagons). 
The current density in the FiM layer was $j_{\textrm{FiM}}\sim \SI{3.5e10}{A/m^2}$ in both measurements. 
}
\end{figure}

We fabricated three different sets of samples, all grown on SiO\textsubscript{2} substrates by magnetron sputtering: (i) the full heterostructures Ni$_x$Cu$_{1-x}$(\SI{6}{\nano\meter})/Ti(\SI{1}{\nano\meter})/FiM/Si$_3$N$_4$ ($x = 40,\,50,\,55,\,70\%$), referred to as Ni$x$, where FiM stands for the ferrimagnetic multilayer [Gd(\SI{0.4}{\nano\meter})/Fe(\SI{0.27}{\nano\meter})]$_{30}$, (ii) the reference layers Ni$_x$Cu$_{1-x}$(\SI{6}{\nano\meter})/Si$_3$N$_4$, referred to as Ni$x-r$, and (iii) the single FiM layer Ti(\SI{1}{\nano\meter})/FiM/Si$_3$N$_4$. The Ni$_x$Cu$_{1-x}$ layers are amorphous, as revealed by X-ray diffraction. The Ni$_x$Cu$_{1-x}$ layers in Ni40, Ni50, and Ni55 are paramagnetic at room temperature, which avoids spurious magnetotransport signals in the electrical measurements of SOTs. The Ti layer in all the heterostructures serves as a spacer to suppress possible magnetic proximity effects between Ni$_x$Cu$_{1-x}$ and the FiM. A sketch of the Ni$x$ samples is shown in Fig.~\ref{fig1}(a).
The samples were patterned in \SI{7.5}{}-\SI{}{\micro\meter}-wide and \SI{11}{}-\SI{}{\micro\meter}-long Hall bar devices using photolithography and lift-off, as shown in Fig.~\ref{fig1}(b).

Figure~\ref{fig1}(c) shows $M_\textrm{s}\,t$, the magnetic moment per unit area, as a function of out-of-plane magnetic field of Ni$55$, its reference Ni$55-r$, and the single FiM sample. Here, $M_\textrm{s}$ represents the saturation magnetization, $t$ the thickness of the magnetic layer, and $M_\textrm{s}\,t$ is the total magnetic moment of the as-grown layers measured at room temperature by a superconducting quantum interference device divided by the sample area. 
The comparison between the different curves shows that Ni$55-r$ develops a sizable magnetization in a magnetic field, which adds to that of the FiM in the Ni$55$ sample.
Because we consider the effect of the SOTs on the sole FiM, we estimate an effective magnetic moment per unit area for each Ni$x$ sample as the difference between the moments of the full heterostructures Ni$x$ and their references Ni$x-r$, namely $M_\textrm{s}\,t = (M_\textrm{s}\,t)_{\textrm{Ni}x} - (M_{\textrm{s}}\, t)_{{\textrm{Ni}x-r}}$.
The magnetotransport measurements were performed by injecting an alternate current of frequency $\omega/(2\pi) = \SI{10}{\hertz}$ and density $j$ and computing the respective Hall ($R^{\textrm{H}}_{1,2\omega}$) and longitudinal ($R^{\textrm{L}}_{1,2\omega}$) components of the first and second harmonic resistances from the measured harmonic voltages.\cite{GarelloNatNanotec2013, AvciPRB2014} 
The first harmonic Hall resistance accounts for the contributions from the anomalous Hall effect (AHE) and planar Hall effect (PHE) and is given by
\begin{equation}
    R^{\textrm{H}}_{1\omega} = R_{\textrm{AHE}}\cos\vartheta + R_{\textrm{PHE}}\sin^2\vartheta\sin2\varphi,
\end{equation}
where $R_{\textrm{AHE}}$ and $R_{\textrm{PHE}}$ are the anomalous and planar Hall resistance coefficients, and $\vartheta$ and $\varphi$ are the polar and azimuthal angles between the magnetization and the $z$- and $x$-axis, respectively. 
Figure~\ref{fig1}(d) shows $R^{\textrm{H}}_{1\omega}$ as a function of out-of-plane field for Ni55 and the single FiM sample.
The signal is dominated by the AHE of the FiM in both samples; the remanence and polarity of the curves indicate that the FiM has perpendicular magnetic anisotropy and its magnetization is Fe-like. The amplitude of the AHE in Ni55 is reduced compared to the single FiM sample because part of the current flows in the paramagnetic Ni$_x$Cu$_{1-x}$ layer and Ni has opposite sign of $R_{\textrm{AHE}}$ compared to Fe. \cite{OmoriPRB2019} Moreover, the magnetic properties and AHE of FiM layers are known to depend on the adjacent layers, which are different in the two samples.\cite{HaltzPhysRevMater2018, ZhangPRB2020, DangPRB2020, BelloAPL2022}

\begin{figure}[t!]
\centering
\includegraphics[width=8.5cm,keepaspectratio]{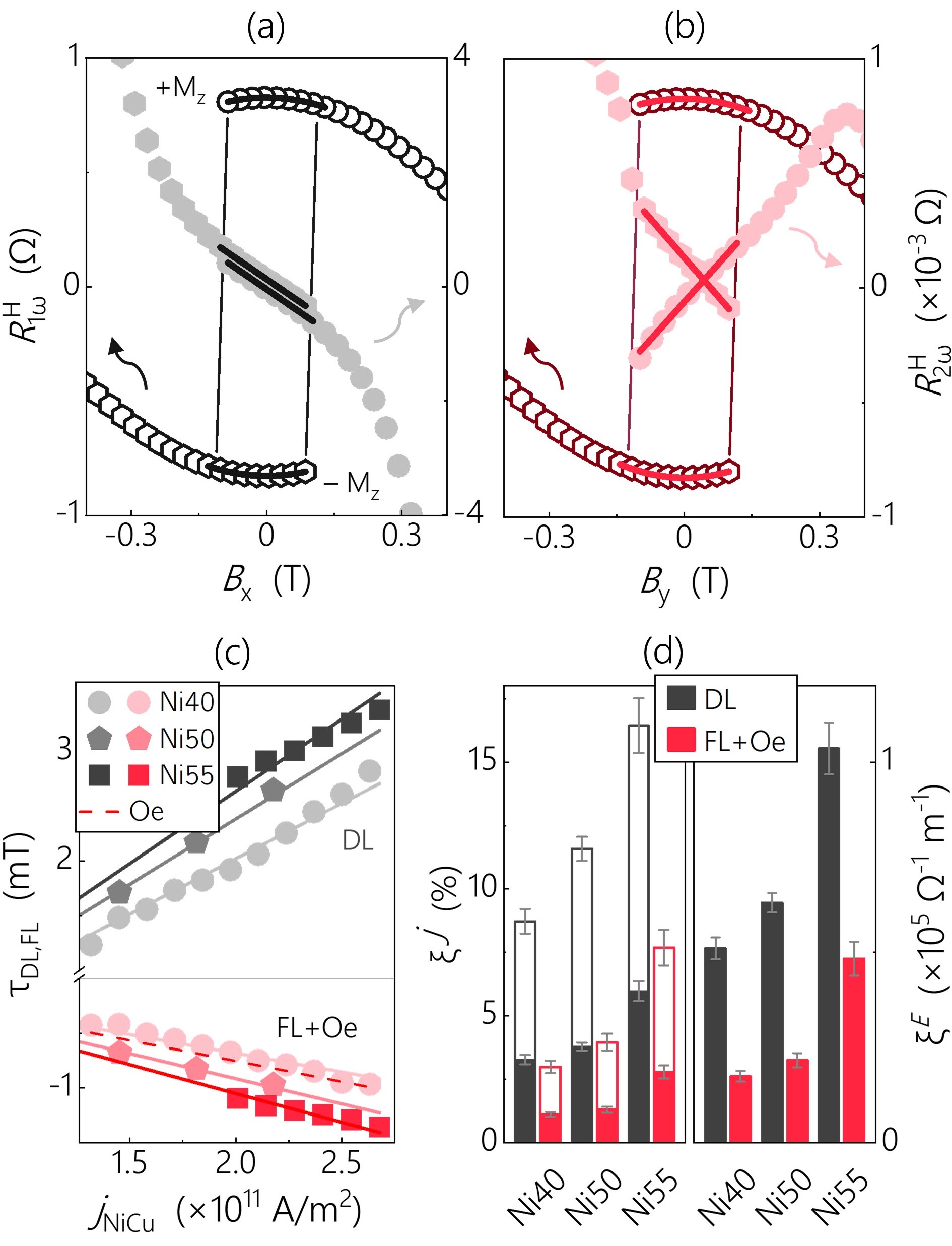}
\caption{\label{fig2} 
(a) and (b) $R^{\textrm{H}}_{1\omega}$ (left axis) and $R^{\textrm{H}}_{2\omega}$ (right axis) of Ni40 as a function of external magnetic field applied at an angle $\vartheta_\textrm{B} = 87^{\circ}$ along (a) the $x$-axis (left plot) and (b) the $y$-axis (right plot). The solid lines are fits used to estimate $\partial^2 R^{\textrm{H}}_{1\omega}/\partial^2 H$ and  $\partial R^{\textrm{H}}_{2\omega}/\partial H$ in Eq.~\ref{torques}. 
(c) DL (black) and FL (red) effective fields of Ni40, Ni50, and Ni55 as a function of the current density in the Ni$_x$Cu$_{1-x}$ layer. The solid lines are linear fits with intercept forced to zero. The error bars, which include the fit errors in (a) and (b) and the uncertainties in $R_{\textrm{AHE}}$ and $R_{\textrm{PHE}}$, are smaller than the symbols. 
(d) DL (black) and FL (red) SOT efficiencies of Ni$x$. 
The left panel shows the SOT efficiencies with respect to the total current density $\upxi^j$ (empty columns) and to the current density in the sole Ni$_x$Cu$_{1-x}$ layer, $\upxi^{j_\textrm{{NiCu}}}$ (full columns); the right panel shows the SOT efficiencies with respect to the electric field $\upxi^E$.  The errors are calculated from the uncertainty of the fits in (c) and $M_\textrm{s}\,t$.}
\end{figure}

The SOTs were measured using the harmonic Hall voltage method, which relies on $R^{\textrm{H}}_{2\omega}$ as a measure of the current-induced oscillations of the magnetization.\cite{GarelloNatNanotec2013, HayashiPRB2014} The SOTs can be decomposed into two components orthogonal to the unitary magnetization \textbf{m}, to define a damping-like (DL) torque, $\textbf{T}_\text{DL} = \tau_\text{DL}\,\textbf{m}  \times (\boldsymbol{\zeta} \times \textbf{m})$, and a field-like (FL) torque, $\textbf{T}_\text{FL} = \tau_\text{FL}\, \textbf{m} \times \boldsymbol{\zeta}$. Here, the torques are normalized by the magnetization, $\boldsymbol{\zeta}$ is the unit vector representing the spin polarization of the injected spin current, and $\tau_\text{DL}$ and $\tau_\text{FL}$ are the amplitudes of the effective magnetic fields describing the action of the SOTs on the magnetization. We estimated $\tau_\text{DL}$ and $\tau_\text{FL+Oe}$, with Oe denoting the Oersted field, using the following equations valid under the assumption $\vartheta \to 0$ \cite{HayashiPRB2014}
\begin{align}
    \tau_{\textrm{DL}} = 2\frac{B_x \pm 2\chi B_y}{1-4\chi^2},   \\
    \tau_{\textrm{FL+Oe}} = 2\frac{B_y \pm 2\chi B_x}{1-4\chi^2},  
    \label{torques}
\end{align}
where $\chi = R_{\textrm{PHE}}/R_{\textrm{AHE}}$ ($\sim 10^{-2}$ in our samples)  and $B_{x,y} \equiv \bigl(\frac{\partial R^{\textrm{H}}_{2\omega}}{\partial H}/\frac{\partial^2 R^{\textrm{H}}_{1\omega}}{\partial^2 H}\bigr)\big|_{\hat{m}\parallel {x,y\,}}$. As an illustrative case, Figs.~\ref{fig2}(a) and (b) show $R^{\textrm{H}}_{1,2\omega}$ of Ni40 as a function of a nearly in-plane external magnetic field oriented along the $x$- and $y$-axes measured at $j = \SI{7.4e10}{\A/\m^2}$ ($j$\textsubscript{NiCu} $= \SI{1.9e11}{A/m^2}$) and the fits (solid lines) used to estimate the SOT effective fields. Figure~\ref{fig2}(c) shows $\tau_{\textrm{DL}}$ and $\tau_{\textrm{FL+Oe}}$ vs $j_\text{NiCu}$ for the different Ni$x$ samples. Note that the total FL+Oe field is similar to the expected Oersted field from Biot-Savart's law, evidencing the small contribution of the FL torque in our samples.

The SOT efficiencies with respect to the applied current density ($j$) and electric field driving the current ($E$) are given by\cite{ManchonRMP2019, NguyenPRL2016}
\begin{align}
    \upxi^{j}_{\textrm{DL,FL}} &= \frac{2e}{\hbar}M_{\textrm{s}}t\frac{\tau_{\textrm{DL,FL}}}{j}, \label{eq:efficiencyJ}\\
    \upxi^{E}_{\textrm{DL,FL}} &= \frac{2e}{\hbar}M_{\textrm{s}}t\frac{\tau_{\textrm{DL,FL}}}{E},
    \label{eq:efficiencyE}
\end{align}
where $e$ is the elementary charge and $\hbar$ is the reduced Planck constant. 
$\upxi^j$ ($\upxi^E$) represents the ratio of the effective spin current absorbed by the magnetic layer relative to the applied charge current (electric field) and can be thus regarded as an effective spin Hall angle (effective spin Hall conductivity) for a specific combination of a spin injector and magnetic layer. Because distinguishing between SOTs generated by spin and orbital currents is not straightforward,\cite{SalaPRR2022} $\upxi^j$ and $\upxi^E$ include also the effect of orbital currents. The efficiencies calculated by Eqs.~\ref{eq:efficiencyJ} and \ref{eq:efficiencyE} constitute a lower bound for quantifying the charge-to-spin conversion of a spin injector since they do not account for the spin transparency of the interface between the spin injector and the magnetic material. 

Figure~\ref{fig2}(d) presents the SOT efficiencies calculated by averaging over two devices for each Ni$x$ sample. The sign of the DL torque is consistent with that of Pt/Co, considering the same stacking order between the spin injector and the magnetic layer. Both $\upxi_{\textrm{DL}}$ and $\upxi_{\textrm{FL+Oe}}$ increase monotonically with Ni content and reach a maximum at $x = 55\%$ in the measured composition range: $\upxi^{j}_{\textrm{DL}} = 0.16$ ($\upxi^{E}_{\textrm{DL}} = 1.04\times10^5~\SI{}{\ohm^{-1}\meter^{-1}}$) and $\upxi^{j}_{\textrm{FL+Oe}} = 0.08$ ($\upxi^{E}_{\textrm{FL+Oe}} = 0.48 \times10^5~\SI{}{\ohm^{-1}\meter^{-1}}$) at $x = 55\%$ are comparable with the values reported for 5$d$ metal system such as Pt/Co.\cite{ManchonRMP2019}
In making such a comparison, one should note that the effective spin Hall angle changes if it is calculated with respect to the average current density ($\upxi^{j}$) or to the current density flowing in the spin injector ($\upxi^{j_\textrm{NiCu}}$) estimated using a parallel resistor model. In the latter case, we obtain $\upxi^{j_\textrm{NiCu}}_{\textrm{DL}} = 0.06$ and $\upxi^{j_\textrm{NiCu}}_{\textrm{FL+Oe}} = 0.03$, which are about half of the values reported for Pt/Co. The resistivity and magnetization parameters used to calculate the different $\upxi$ are reported in Table~\ref{tab1}.

We now compare the SOT efficiencies of the Ni$x$ samples with the measurements of spin-charge interconversion reported for Ni$_x$Cu$_{1-x}$ alloys in the literature. 
Keller \emph{et al.} \cite{KellerPRB2019} reported $\upxi^E_{\textrm{DL}}\sim1\times10^5~\SI{}{\ohm^{-1}\meter^{-1}}$ at $x = 60\%$, before considering any correction due to interfacial spin memory loss. This efficiency, normalized by the reported electrical conductivity of the NiCu layer of $\SI{1.93e6}{\ohm^{-1}\meter^{-1}}$, corresponds to a spin Hall angle of 0.05.
Varotto \emph{et al.} \cite{VarottoPRL2020} reported a spin Hall angle of $0.04$ at $x = 60\%$, in line with our results. Because the spin diffusion length $\lambda_s$ decreases with increasing Ni content \cite{HsuPRB1996} and the measured values of $\tau_\text{DL}$ and $\tau_\text{FL}$ depend on $\lambda_s$ if $t_{\rm NiCu} \lesssim 2\lambda_s$, an alternative figure of merit is the product $\upxi^{j_\textrm{NiCu}}\lambda_s$. Assuming $\lambda_s \sim \SI{4}{\nano\meter}$ at $x = 40\%$ and \SI{3}{\nano\meter} at $x = 55\%$,\cite{HsuPRB1996} we obtain $\upxi_\textrm{DL}^{j_\textrm{NiCu}} \lambda_s = \SI{0.12}{\nano\meter}$ for Ni40 and \SI{0.18}{\nano\meter} for Ni55. These values are similar to $\upxi_\textrm{DL}\lambda_s = \SI{0.1}{\nano\meter}$ obtained in Ref.~\onlinecite{VarottoPRL2020}. Overall, this comparison shows that Ni$_x$Cu$_{1-x}$ alloys present large and comparable figures of merit for charge-to-spin and spin-to-charge conversion. 

\begin{figure}[t!]
\centering
\includegraphics[width=8.5cm,keepaspectratio]{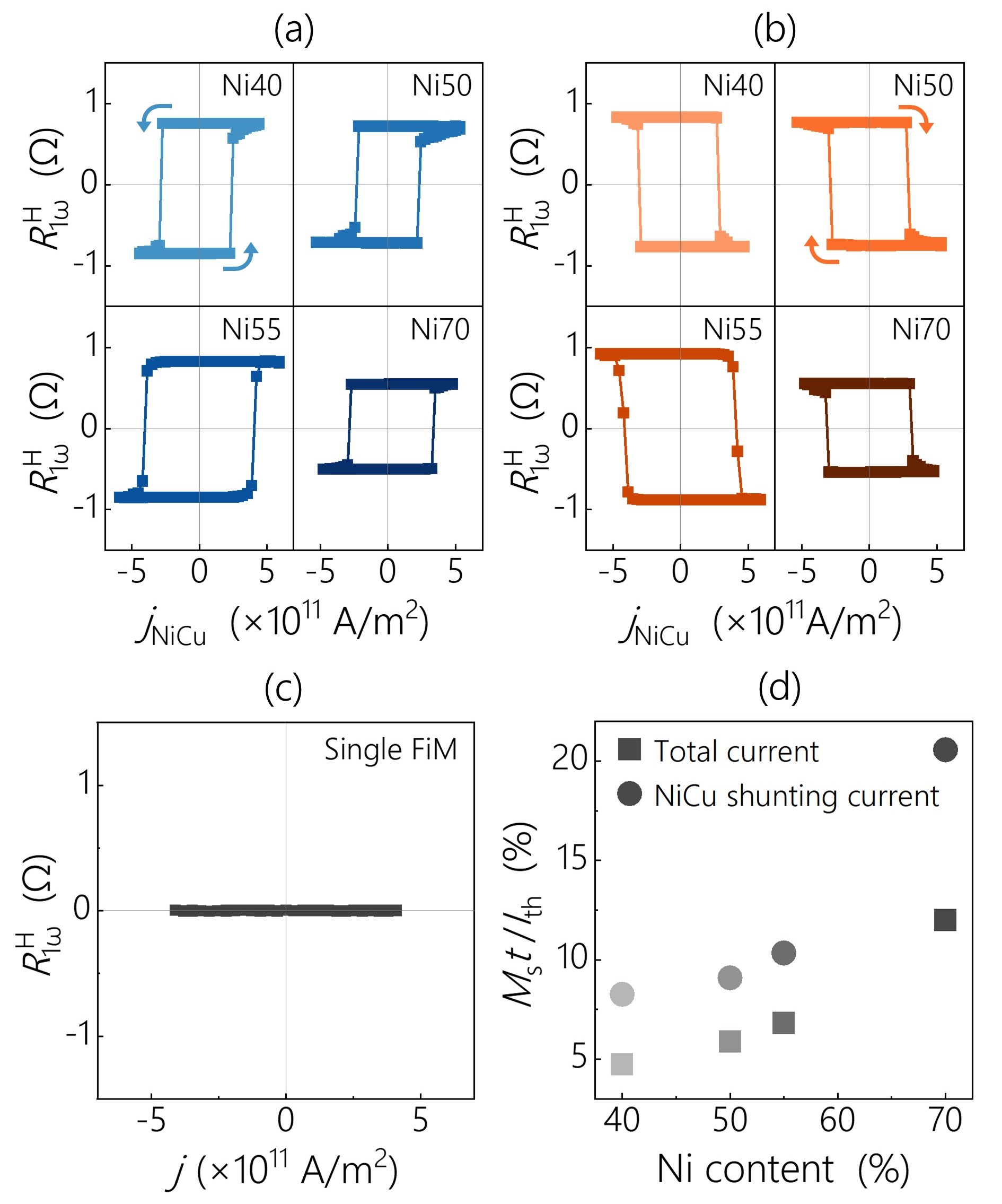}
\caption{\label{fig3} SOT-induced magnetization switching. (a) and (b) Anomalous Hall resistance as a function of current density in Ni$_x$Cu$_{1-x}$. The switching loops are acquired using $\SI{10}{}$-$\SI{}{\ms}$-long current pulses in the presence of an in-plane magnetic field of \SI{10}{\milli\tesla} oriented along (a) $-x$ and (b) $+x$.
(c) Anomalous Hall resistance as a function of current density for the single FiM sample. The loop is acquired using $\SI{10}{}$-$\SI{}{\ms}$-long current pulses and an in-plane magnetic field of \SI{170}{\milli\tesla} oriented along $+x$. (d) Ratio of the magnetic moment per unit area to the threshold switching current as a function of Ni content.} 
\end{figure}

\begin{table}[h!]
\centering
\footnotesize
\setlength{\tabcolsep}{4pt}
\renewcommand{\arraystretch}{1.5}
\begin{tabular}{llcccccc}
\hline\hline
\multicolumn{1}{l}{\makecell{\\Sample}} & \makecell{$M_{\textrm{s}}\,t$\\($\SI{}{\milli\ampere})$} & \multicolumn{1}{l}{\makecell{$R_{\textrm{AHE}}$\\$(\SI{}{\ohm})$}} & \makecell{$\rho$\\$(\SI{}{\micro\ohm\cm})$} & \multicolumn{1}{l}{\makecell{$\rho$\textsubscript{NiCu}\\$(\SI{}{\micro\ohm\cm})$}} & \makecell{$\rho$\textsubscript{FiM}\\$(\SI{}{\micro\ohm\cm})$} & \multicolumn{1}{l}{\makecell{$R/R\textsubscript{NiCu}$\\$(\%)$}} \\ \hline
Ni40  &  1.0  & 0.8  & 170  & 64  & 320  & 58  \\
Ni50  &  1.1  & 0.7  & 181  & 60  & 423  & 67  \\
Ni55  &  1.7  & 1.0  & 157  & 58  & 308  & 60  \\
Ni70  &  2.9  & 0.5  & 129  & 49  & 242  & 58  \\     
FiM   &  1.3  & 7.5  & 335  & -   & 335  & -   \\  
\hline\hline
\end{tabular}
\caption{Summary of the magnetic and electrical properties of Ni$x$ and the single FiM sample at room temperature. From left to right: $M_{\textrm{s}}\,t$ of the FiM, $R_{\rm AHE}$, resistivity of Ni$x$, Ni$x-r$, and FiM layer, and current branching ratio in the Ni$_x$Cu$_{1-x}$ layer. For the Ni$x$ samples, $\rho_\textrm{FiM}$ was estimated using a parallel resistor model.}
\label{tab1}
\end{table}

Based on the large charge-to-spin conversion efficiency, we further demonstrate SOT-induced magnetization switching of Ni$x$ for four different alloy's compositions ($x = 40,\,50,\,55,\,70\%$). To do so, we injected $\SI{10}{}$-$\SI{}{\ms}$-long current pulses to induce magnetization switching and inspected the magnetization state by reading $R^H_{1\omega}$ using an alternate current of frequency $\omega/(2\pi) = \SI{115}{\hertz}$ after each pulse. As required for samples with perpendicular magnetic anisotropy, we apply an in-plane magnetic field to break the symmetry of the DL torque and define the switching polarity.\cite{MironNat2011} Figures~\ref{fig3}(a) and (b) show the switching loops of Ni$x$ acquired for negative and positive external magnetic fields of amplitude $\SI{10}{\milli\tesla}$ applied along the $x$-axis. For all the samples, we observed full magnetization switching, as demonstrated by the semi-amplitude of the switching loops, matching the saturation $R_\text{AHE}$ values. As expected, the switching polarity changes upon inversion of the in-plane magnetic field. In addition, the switching polarity is the same as observed in Pt/Co,\cite{MironNat2011} consistently with the sign of the SOTs. We further verified that switching occurs reliably using $\SI{100}{}$-$\SI{}{\ns}$-long current pulses and that the threshold switching current $I_\textrm{th}$ decreases monotonically with the applied in-plane field. Measurements of the single FiM sample gave null results, as shown in Fig.~\ref{fig3}(c), excluding the possibility of switching induced by self-torques in the FiM material.\cite{CespedesAdvMater2021}

$I_\textrm{th}$ is generally influenced by the coercivity and magnetic anisotropy of the FiM layer, which can vary from sample to sample. If such variations are neglected, the ratio $M_\text{s}\,t/I_\textrm{th}$ can be taken as a rough measure of the SOT switching efficiency. Figure~\ref{fig3}(d) shows that $M_\text{s}\,t/I_\textrm{th}$ measured in the different Ni$x$ samples improves with increasing Ni content, in agreement with the trend of the SOT efficiency reported in Fig.~\ref{fig2}(d). The same trend is observed if $I_\textrm{th}$ is taken as the total current (squares) or only the current flowing in the Ni$_{x}$Cu$_{1-x}$ layer (dots). The improved switching efficiency at $x = 70\%$ is also in agreement with the higher charge-to-spin conversion efficiency measured by ferromagnetic resonance.\cite{KellerPRB2019}

In conclusion, our results show that Ni$_x$Cu$_{1-x}$ alloys are very efficient generators of SOTs. Current injection in Ni$_x$Cu$_{1-x}$ is further shown to induce reliable switching of the magnetization of a 20-nm-thick [Gd/Fe] multilayer at a current density of about $5\times 10^{11}$~A/m$^2$. The SOT efficiency increases with Ni content, reaching a maximum spin Hall conductivity of $\upxi^{E}_{\textrm{DL}} = 1.04\times10^5~\SI{}{\ohm^{-1}\meter^{-1}}$ in Ni55, which is comparable to that of heavy metal systems.   

\begin{acknowledgments}
This work was supported by the Swiss National Science Foundation (Grant No. 200020-200465). P.N. acknowledges the support of the ETH Zurich Postdoctoral Fellowship Program 19-2 FEL-61. We thank Laura van Schie for her assistance in performing switching measurements using $\SI{100}{}$-$\SI{}{\ns}$-long current pulses.
\end{acknowledgments}

The authors have no conflicts to disclose.

The data that support the findings of this study are openly available from the ETH Research Collection at https://doi.org/10.3929/ethz-b-000633473.

\nocite{*}
\bibliography{aipsamp}

\end{document}